\documentclass[12pt,a4paper]{article}

\usepackage{amsfonts}
\usepackage{amssymb,amsmath}
\usepackage{graphicx}

\newcommand{\be}{\begin{equation}}
\newcommand{\ee}{\end{equation}}
\newcommand{\bea}{\begin{eqnarray}}
\newcommand{\eea}{\end{eqnarray}}
\newcommand{\la}{\langle}
\newcommand{\ra}{\rangle}

\newcommand{\ld}{\left(}
\newcommand{\rd}{\right)}
\newcommand{\lb}{\left\{}
\newcommand{\rb}{\right\}}
\newcommand{\lbr}{\left[}
\newcommand{\rbr}{\right]}

\newcommand{\ve}{\varepsilon}
\newcommand{\bm}{\boldsymbol}

\begin{document}

\title{Spatial decay of the one-body density matrix\\
in insulators revised}

\author{Janusz J\c{e}drzejewski$^{1,2,\star}$ and
        Taras Krokhmalskii$^{3,1,\ast}$\\
\sf        $^{1}$Institute of Theoretical Physics,
                    University of Wroc\l aw\\
\sf        pl. Maksa Borna 9, 50-204 Wroc\l aw, Poland\\
\sf        $^{2}$Department of Theoretical Physics,
                    University of \L \'{o}d\'{z}\\
\sf        149/153 Pomorska Str., 90-236 \L \'{o}d\'{z}, Poland\\
\sf        $^{3}$Institute for Condensed Matter Physics\\
\sf        1 Svientsitskii Str., L'viv-11, 79011, Ukraine}

\date{\today}

\maketitle

\vspace{-8mm}
\begin{abstract}

In the framework of the band theory, we consider two tight-binding
models of insulators. The first one, proposed recently by Taraskin
et al, is a translationally invariant system, built out of two
independent non-overlapping bands of single-particle orbitals that
are coupled by a weak inter-band hybridization. This kind of
insulator exhibits unphysical properties: we show, in particular,
that the one-body density matrix does not depend on the width of
the gap between the bands. Consequently, there is no
delocalization effect with increasing metallicity. In the second
model there are also two bands. However, they are not imposed by
construction but are created from a band of single-particle orbitals
due to the breaking of the translational symmetry by a periodic
potential. These bands are separated by a gap for all nonzero
values of the unique energy parameter of the model. We demonstrate
that the one-body density matrix has the same structure  as in the
first model. As a result, the large distance asymptotic formulae
derived by Taraskin et al in dimensions $D=1,2,3$, apply as well,
but only for very large gap widths. In $D=1$ and in the diagonal
direction of $D=2$ cases, we derive a stronger asymptotic formula,
valid for all gap widths. The both kinds of asymptotic formulae
are composed of a dimension-dependent power-law factor and a
gap-dependent exponentially decaying factor. The latter
asymptotics implies that the exponential decay rate
vanishes linearly with the vanishing gap. In non-diagonal directions,
we have found numerically that the linear scaling is replaced by the
square root one. Independently of the direction, the exponential decay
rate grows logarithmically with the gap width, for sufficiently large
gaps.

\end{abstract}

\vspace{1mm}

\vspace{1mm}


\vspace{5mm}


\noindent
E-mail addresses: \\
$\star$ --- jjed@ift.uni.wroc.pl \\
$\ast$ --- krokhm@icmp.lviv.ua\\

\newpage

The rapid progress in computational techniques used for
calculating properties of solids enabled researchers to use a
localized real-space approach to describing such properties. It
has made possible large-scale calculations based on the density
functional theory \cite{payne}, in particular calculating  the
electronic structure of solids by means of $O(N)$ methods
\cite{goed,kohn1}. In all these methods it is the one-body density
matrix (DM) that is the central quantity. Its decay rate
determines the degree of locality of all relevant quantities and
is decisive for the speed of the involved algorithms. This is why
one observes a growing interest in localization properties of DM
in recent years. However, the first result concerned with the rate
of decay of DM was published by W. Kohn as early as in 1959
\cite{kohn2}. It has been established that in a 1D model of an
insulator the DM decays exponentially with distance. The
exponential decay of DM was reconsidered by Ismail-Beigi and Arias
\cite{beigi} a few years ago (see also the references quoted there
for developments in the meantime) in a general, model-independent
context, for systems described by single-particle orbitals in
periodic potentials in arbitrary dimensions. Then, He and
Vanderbilt \cite{he} have discovered  the power-law prefactor
mutiplying the exponential factor in the asymptotic formula for
DM. Finally, Taraskin et al \cite{taraskin1} presented their
minimal model of an insulator and described, analytically and
numerically a power-law decay and an exponential
decay in dimensions $D= 1,2,3$.

In this communication we reconsider the model of insulator
proposed in \cite{taraskin1}. This is a translationally invariant
system described by two kinds of single-particle orbitals between
which the electrons can hop. We demonstrate that it is not a kind
of insulator considered in \cite{beigi} or earlier papers on the
subject. For instance, for those values of the energy
parameters of the model for which there exist two bands separated
by a gap, there is no relation between the decay rate of DM and
the size of the gap. We have succeeded in deriving analytically
the large distance properties of DM of another system, described
by one kind of single-particle orbitals and a periodic external
potential. This system exhibits two bands separated by a gap for
all nonzero values of the parameter determining the strength of
the potential. Specifically, in the large distance behavior of
DM elements of this system we have determined the power-law factor
and the exponential factor. We have found also the scaling law of
the inverse correlation length versus the vanishing gap.

The model of an insulator proposed in \cite{taraskin1} (TDE
insulator) is described by the following second-quantized
Hamiltonian

\be
\label{hTDE}
H_{tde}=\sum_{\bm i;\,\mu}\ve_{\mu}a^+_{\bm
i,\,\mu}a_{\bm i,\,\mu} +\sum_{\la \bm i,\,\bm j \ra;\, \mu,\,\nu}
t_{\mu \nu}\ld a^+_{\bm i,\,\mu}a_{\bm j,\,\nu}+h.c.  \rd .
\ee

\begin{figure}
\begin{center}
\includegraphics[clip=on,width=8cm]{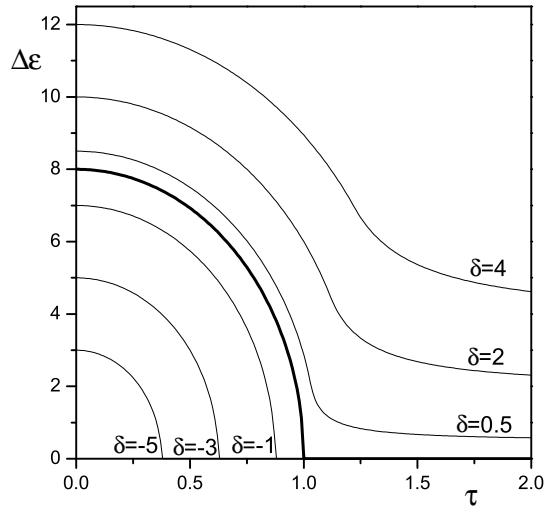}
\end{center}
\caption[]{Curves  of constant gap $\delta$ for $D=2$.
The thick curve corresponds to $\delta =0$;
it splits the $\tau \geq 0$, $\Delta\ve \geq 0$ quadrant into two
disjoint regions of $\delta \geq 0$ and $\delta < 0$.}
\label{fig1}
\end{figure}

In the above expression, $\bm i,\bm j$ represent the lattice sites
of a $D$-dimensional Bravais lattice, while $\la \bm i,\bm j \ra$
stands for a pair of nearest neighbors on this lattice. The
operators $\{a^{+}_{\bm i,\mu},a_{\bm i,\mu} \}$ create,
annihilate, respectively, a spinless fermion in a single-particle
orbital $|\bm i,\mu \ra $ whose bare energy is $\ve_{\mu}$, where
$\mu$ differentiates between the two kinds of involved orbitals.
The orbitals $|\bm i,\mu \ra$ satisfy the orthonormality condition
$\la \bm j,\nu|\bm i,\mu \ra= \delta_{\bm i,\,\bm
j}\delta_{\nu,\,\mu}$. The hopping integrals between the orbitals
$|\bm i,\mu \ra $ and $|\bm j,\nu \ra$ are $t_{\mu\nu}$, and they
are nonzero only if $\bm i,\bm j$ are nearest neighbors. If the
periodic boundary conditions are imposed, then the system is
diagonalized by passing to plane-wave orbitals $|\bm k,\mu \ra $,
with the wave vector $\bm k$ in the first Brillouin zone. In order
to reduce the number of energy parameters in the model, we make
the bands symmetric. Specifically, we equalize the intraband
transfer integrals $t_{11}=t_{22}=t$, introduce $\Delta\ve \equiv
\ve_1-\ve_2$, with $t > 0$ and $\Delta\ve>0$ for definitness, and
shift also the zero of the energy scale to $\ld \ve_1+\ve_2
\rd/2$. Moreover, we express all the energies in the units of the
intraband hopping $t$. Then, the eigenenergies of the system, as
functions of the wave vector $\bm k$, form the upper,
$\lambda^{+}_{\bm k}$, and the lower, $\lambda^{-}_{\bm k}$, bands
$\lambda^{\pm}_{\bm k}$: \be \label{TDEdisp} \lambda^{\pm}_{\bm k}
= 2S_{\bm k} \pm 2 \lbr \ld{\Delta\ve}/{4}\rd^2 + \tau^2S^2_{\bm
k} \rbr^{1/2}, \ee where $\tau \equiv t_{12}/t$ is the interband
transfer integral in the units of $t$, $\tau > 0$ for definitness,
and $S_{\bm k}=\frac{1}{2} \sum_{\bm j;\, \la \bm i,\,\bm j \ra}
\exp\ld i{\bm k(\bm i\!-\! \bm j)}\rd$ stands for the structure
factor. Clearly, $\lambda^{+}_{\bm k}$ and $\lambda^{-}_{\bm k}$
are mutually symmetric about zero and there are two independent
energy parameters: $\tau$ and $\Delta\ve$. Let $\delta \equiv
min_{\bm k}\lambda^{+}_{\bm k} - max_{\bm k} \lambda^{-}_{\bm k}$
be the width of the gap (if $\delta \geq 0$) or the overlap of the
two bands (if $\delta < 0$). In Fig.~\ref{fig1}, we display the
curves of constant $\delta$, in the plane $(\tau, \Delta\ve$), for
$D=2$ . It is seen that if the bare energies of the two kinds of
single-particle orbitals are such that $\Delta\ve > 8$, then for
any hybridization $\tau > 0$ the width of the gap $\delta > 0$.
Otherwise, the hybridization has to be large enough. Generally, we
need $\Delta\ve > 4D$ to have a gap for arbitrarily small $\tau$.
Now, consider the TDE insulator, i.e. we set $\delta > 0$ and fill
up completely with spinless electrons the lower band. Moreover,
the underlying Bravais lattice is assumed to be a $D$-dimensional
simple cubic lattice, with lattice vectors expressed in the units
of the lattice constant (i.e. the components $j_l$, $l=1,\ldots,
D$, of lattice vectors $\bm j$ are integer). Then, the
zero-temperature matrix elements of DM are given by the
zero-temperature averages $\la a^+_{\bm i,\,\mu} a_{\bm i + \bm r,\,\nu}
\ra$, which in the thermodynamic limit assume the form
\be
\label{DM11,22}
\la a^+_{\bm i,\,{\substack{1\\2}}}
a_{\bm i + \bm r,\,{\substack{1\\2}}} \ra=
(2(2\pi)^{D})^{-1}
\int_{-\pi}^{\pi}\!\!\!\!\!\cdots\!\!\!\int_{-\pi}^{\pi}
{\mbox d}\bm k  \exp\ld i \bm k \bm r \rd
\lbr 1 \mp A \Delta^{-1} (\bm k) \rbr ,
\ee

\be
\label{DM12}
\la a^+_{\bm i,\,1}a_{\bm i + \bm r,\,2} \ra=
-(2(2\pi)^{D})^{-1}
\int_{-\pi}^{\pi}\!\!\!\!\!\cdots\!\!\!\int_{-\pi}^{\pi}
{\mbox d}\bm k \exp\ld i {\bm k \bm r} \rd S_{\bm k}\Delta^{-1} (\bm k),
\ee
where $A\equiv \Delta\ve/4\tau$ and
$\Delta (\bm k) \equiv\ld A^2 + S^2_{\bm k} \rd^{1/2}$.
Clearly, the matrix elements of DM depend on the two energy parameters
of TDE insulator only through their ratio $\Delta\ve / \tau$, hence
in that part of $(\tau, \Delta\ve$)-plane where $\delta \geq 0$, they are
constant along the rays with the slope $4A$, emerging from the origin
$(0,0)$. Consequently, the DM does not depend on the width of the gap
between the upper band and the lower one. In other words, by following a
suitable curve in the region $\delta \geq 0$, we can impose any relation
between the parameter $A$ and the gap width $\delta$. For instance,
along the lines of constant $\tau$ the parameter $A$ is an increasing
function of $\delta$, whose minimum is $A_\tau$, and the limit
$A \rightarrow A_\tau^+$, is equivalent to the limit
$\delta  \rightarrow 0^+$. Only for a sufficiently strong hybridization,
$\tau \geq 1$, the minimum $A_\tau$ attains zero and then
$A \rightarrow 0^+$ is equivalent to $\delta  \rightarrow 0^+$.
The latter limit has been studied numerically in \cite{taraskin1}
in order to obtain the large distance behavior of DM for
$\delta  \rightarrow 0^+$.

Moreover, despite the hybridization of the two types of bands, in
some lattice directions there are no density-density correlations
between the electrons,
that is the correlations $\la a^+_{\bm j,\,\mu} a_{\bm j,\,\mu'}
\ra$ vanish identically in some lattice directions.

Now, consider the model, where only one sort ($\mu$ fixed) of single-particle
orbitals $|\bm j,\, \mu \ra \equiv |\bm j \ra$ is present, whose Hamiltonian is
\be
\label{hcb}
H_{cb}=\sum_{\bm i} U_{\bm i}a^+_{\bm i}a_{\bm i} +
t \sum_{\la \bm i,\,\bm j \ra}
\ld a^+_{\bm i}a_{\bm j}+h.c.  \rd ,
\ee
where we used the same notation as in (\ref{hTDE}) but with the fixed
index $\mu$ suppressed, and $U_{\bm j}$ denotes a periodic external
potential.
Suppose that the underlying Bravais lattice consists of two
interpenetrating sublattices that differ by a primitive translation,
i.e. the nearest neighbors of a site on one sublattice belong
the other one. Then, we set $U_{\bm j}= U_1$ on
one sublattice and $U_{\bm j}=U_2$ on the other one. Under periodic
boundary conditions the Hamiltonian (\ref{hcb}) is diagonalized by plane
wave orbitals $|\bm k \ra$ with the wave vector in the first Brillouin
zone of a sublattice. Specifically, shifting the energy scale to
$(U_1 + U_2)/2$ and expressing all the energies in the units of the
transfer integral $t$, we obtain the upper $\Lambda^+_{\bm k}$  and
the lower $\Lambda^-_{\bm k}$ bands of eigenenergies labeled by ${\bm k}$:
\be
\label{cbdisp}
\Lambda^{\pm}_{\bm k}=\pm  2\sqrt{(u/2)^2+S_{\bm k}^2}
\equiv \pm 2\Delta_{cb}(\bm k ),
\ee
where $u \equiv(U_2-U_1)/2t$ is the unique energy parameter of the model.
Without any loss of generality we can set $u > 0$.
As in the TDE insulator, the two bands are mutually symmetric about zero
and are separated by a nonzero gap $\delta_{cb}=2u$, for any nonzero $u$.
For $u \neq 0$ and completely filled lower band, the system given by
(\ref{hcb}) is an insulator in the sense of the band theory, which we call
the {\it chessboard insulator}. In contrast to TDE insulator, the
chessboard insulator belongs to the class of insulators,
for which a definite scaling of the exponential decay rate versus the
vanishing gap is expected \cite{beigi}.
Let the underlying lattice be
a D-dimensional simple cubic lattice. For definiteness,
we set $U_{\bm j}= U_2$ at the sites of the even sublattice.
The zero temperature, non-diagonal elements of DM are given by
\bea
\label{DM}
\la a^+_{\bm i}a_{\bm i+\bm r} \ra =
(2\pi)^{-D}
\int_{B.Z.}{\mbox d}\bm k \exp\ld i \bm k \bm r \rd
\gamma^{-2}_{\bm k}\lb u^2+(-1)^{\sigma_{\bm r}}\beta^2_{\bm k}
-(-1)^{\sigma_{\bm i}} \lbr 1+(-1)^{\sigma_{\bm r}} \rbr u
\beta_{\bm k} \rb , \eea where the D-dimensional integral is taken
over the first Brillouin zone of a sublattice, $\bm r \neq 0$, and
the following abbreviations have been used \be \label{beta,gamma}
\sigma_{\bm r}\equiv\sum_{l=1}^D r_l , \qquad \beta_{\bm k}\equiv
2S_{\bm k}+2\Delta_{cb}(\bm k)\ , \qquad \gamma^2_{\bm k} \equiv
4\beta_{\bm k}\Delta_{cb}(\bm k). \ee Without any loss of
generality, we can assume that for all $l=1,\ldots,D$, $r_l >0$.
Then, $\sigma_{\bm r}$ is a (noneuclidean) distance between the
lattice point ${\bm r}$ and the origin. To reveal the structure of
DM given by (\ref{DM}), it is convenient to rewrite it separately
for $\sigma_{\bm r}$ even and odd:
\be
\label{DModd}
\la a^+_{\bm i} a_{\bm i+ \bm r} \ra =
-(2\pi)^{-D}
\int_{B.Z.}{\mbox d}\bm k
\exp\ld i\bm k \bm r\rd S_{\bm k} \Delta^{-1}_{cb}(\bm k),
\qquad {\mbox {if}}\qquad \sigma_{\bm r}=2m+1 ,
\ee

\be
\label{DMeven}
\la a^+_{\bm i}a_{\bm i + \bm r} \ra=
-(-1)^{\sigma_{\bm i}}(2(2\pi)^{D})^{-1}u
\int_{B.Z.}{\mbox d}\bm k
\exp\ld i\bm k \bm r \rd  \Delta^{-1}_{cb}(\bm k),
\qquad {\mbox {if}}\qquad \sigma_{\bm r}=2m .
\ee
It appears that the large distance behavior of DM elements for the chessboard
insulator is determined by the function ${\cal{R}}_{cb}(\bm r)$,
\be
\label{Rcb}
{\cal{R}}_{cb}(\bm r)=
(2\pi)^{-D}
\int_{B.Z.}{\mbox d}\bm k
\exp\ld i\bm k \bm r\rd  \Delta^{-1}_{cb}(\bm k),
\ee
since,
\be
\label{DModd'}
\la a^+_{\bm i}a_{\bm i+ \bm r} \ra =
- \frac{1}{2}\sum_{l=1}^{D} {\cal{S}}_l {\cal{R}}_{cb}(\bm r),
\qquad {\mbox {if}}\qquad \sigma_{\bm r}=2m+1,
\ee

\be
\label{DMeven'}
\la a^+_{\bm i}a_{\bm i + \bm r} \ra =
-\frac{u}{2} (-1)^{\sigma_{\bm i}} {\cal{R}}_{cb}(\bm r),
\qquad {\mbox {if}}\qquad \sigma_{\bm r}=2m,
\ee
where ${\cal{S}}_1 {\cal{R}}_{cb}(\bm r)\equiv
{\cal{R}}_{cb}(r_1 + 1,\ldots,r_D) +
{\cal{R}}_{cb}(r_1 - 1,\ldots,r_D)$, etc.
Note that the matrix elements of DM depend on ${\cal{R}}_{cb}(\bm r)$
evaluated only at the points $\bm r$ with $\sigma_{\bm r}$ even.
Similar relations hold in the case of TDE insulator.
On introducing
\bea
\label{R}
{\cal R}(\bm r)\equiv
(2(2\pi)^{D})^{-1}
\int_{-\pi}^{\pi}\!\!\!\!\!\cdots\!\!\!\int_{-\pi}^{\pi}
{\mbox d}\bm k \exp(i\bm k \bm r) \Delta^{-1}(\bm k),
\eea
we obtain
\be
\label{}
\la a^+_{\bm i,1}a_{\bm i + \bm r,2} \ra
=-\frac{1}{2} \sum_{l=1}^D {\cal{S}}_l{\cal {R}}(\bm r),
\ee
\be
\label{}
\la a^+_{\bm i,{\substack{1\\2}}}a_{\bm i + \bm r,{\substack{1\\2}}} \ra
=\mp A{\cal {R}}(\bm r).
\ee
In what follows, we limit our analysis to the $D=1,2$ cases.
Then, one finds easily that
\bea
\label{TDE-cb}
{\cal R}(\bm r)=
\begin{cases}
0,   & \text {if}\ \  \sigma_{\bm r} \text{ is odd },\\
{\cal {R}}_{cb}(\bm r)\rvert_{\frac{u}{2}=A},
&\text {if} \ \  \sigma_{\bm r} \text{ is even }.
\end{cases}
\eea
Therefore, as in the case of the chessboard insulator, the DM
elements depend only on ${\cal R}(\bm r)$ evaluated at the  points
with $\sigma_{\bm r}$ even.
Consequently, the matrix elements
$\la a^+_{\bm i,\,1}a_{\bm i + \bm r,\,2} \ra$ vanish at the lattice
directions with $\sigma_{\bm r}$  even, while
$\la a^+_{\bm i,\, \mu}a_{\bm i + \bm r,\, \mu} \ra$ vanish at the lattice
directions with $\sigma_{\bm r}$ odd.
Moreover, the large distance behavior of the DM elements of both insulators
under considerations, is given by the function ${\cal
{R}}_{cb}(\bm r)$ restricted to the points with $\sigma_{\bm r}$
even. In all the cases studied here we have found that up to a
coefficient independent of $\sigma_{\bm r}$
\be
\label{asymptot}
{\cal {R}}_{cb}(\bm r) \sim \sigma_{\bm r}^{-\gamma}
\exp(-\sigma_{\bm r} / \xi (\epsilon)),
\ee
for sufficiently large
$\sigma_{\bm r}$, where $\epsilon$ stands for an energy parameter
and $\xi(\epsilon)$ is the correlation length.

Apparently, a straightforward way to obtain a large $\sigma_{\bm r}$
asymptotics of ${\cal {R}}_{cb}(\bm r)$ is to expand $\Delta_{cb}(\bm k)$
in powers of $u^{-2}$, for large $u$, carry out the integrals of products
of cosine functions and to approximate the Euler $\Gamma$-functions, that
arise, by the leading term of the Stirling's asymptotic expansion.
Then, in $D=1$ case we obtain the inverse correlation length
\be
\label{TDEasymp1}
\xi^{-1}_{tde}(u)=\ln u+ u^{-2},
\ee
while in directions of finite nonzero slope $\chi\equiv r_1/r_2$ of
the $D=2$ case,
\be
\label{TDEasymp2}
\xi^{-1}_{tde}(u)=\ln u+ u^{-2}(2+\chi+\chi^{-1})
-\sum_{\alpha=-1,1} (1+\chi^{\alpha})^{-1} \ln(1+\chi^{\alpha}),
\ee
and the exponent $\gamma=D/2$. A similar procedure gives asymptotic
formulae along the axes. The asymptotics  (\ref{asymptot}) with (\ref{TDEasymp1})
or (\ref{TDEasymp2}) -- the TDE asymptotics, on setting $u=2A$ amounts to the
corresponding asymptotics derived in \cite{taraskin1}.

\begin{figure}
\begin{center}
\includegraphics[clip=on,width=15cm]{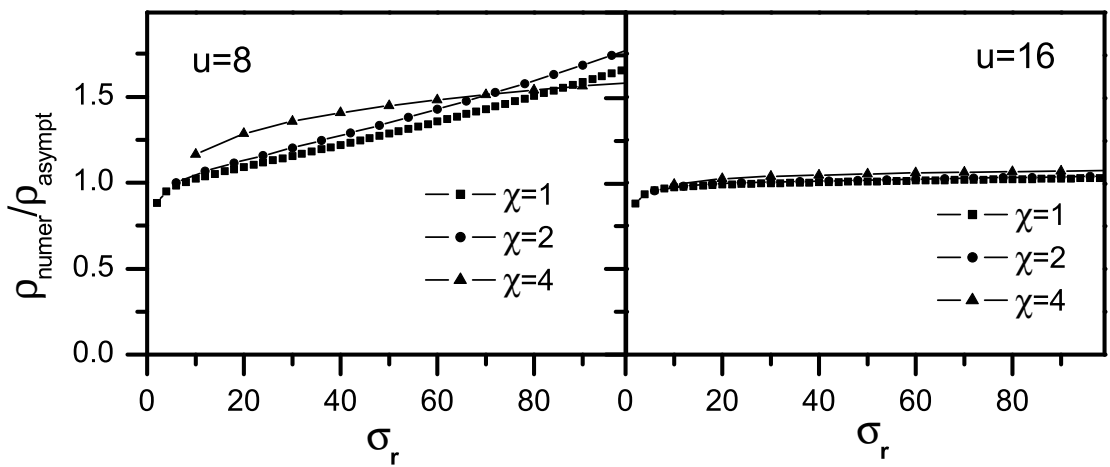}
\end{center}
\caption[]{The calculated numerically exactly function
${\cal {R}}_{cb}(\bm r)$, denoted $\rho_{numer}$, divided by its
TDE asymptotics, denoted $\rho_{asympt}$, which up to a constant
is given by (\ref{TDEasymp2}),
versus distance $\sigma_{\bm r}$, for $D=2$, $\chi =1,2,4$,
and two values of $u$.}
\label{fig2}
\end{figure}

While the TDE asymptotics obtained for the chessboard insulator and the
coresponding one for the TDE insulator have the same structure,
their physical content is very different. The energy parameter $u$ is
in one to one correspondence with the gap width of the chessboard insulator.
In contrast, $A$ is in no definite relation with the gap width of the TDE
insulator.

To test the quality of the asymptotic formula (\ref{asymptot}) with
(\ref{TDEasymp2}), one can compare it with a calculated numerically exactly
function ${\cal {R}}_{cb}(\bm r)$.
The log-log plot of both functions is to rough for such a test.
We have plotted, in Fig.~\ref{fig2},
${\cal {R}}_{cb}(\bm r)$  calculated numerically exactly, divided by its
TDE asymptotics. It is seen that, in the case of D=2 chessboard insulator,
the TDE asymptotics is reasonably good only for as large values of $u$ as 16
or higher.

\begin{figure}
\begin{center}
\includegraphics[clip=on,width=12cm]{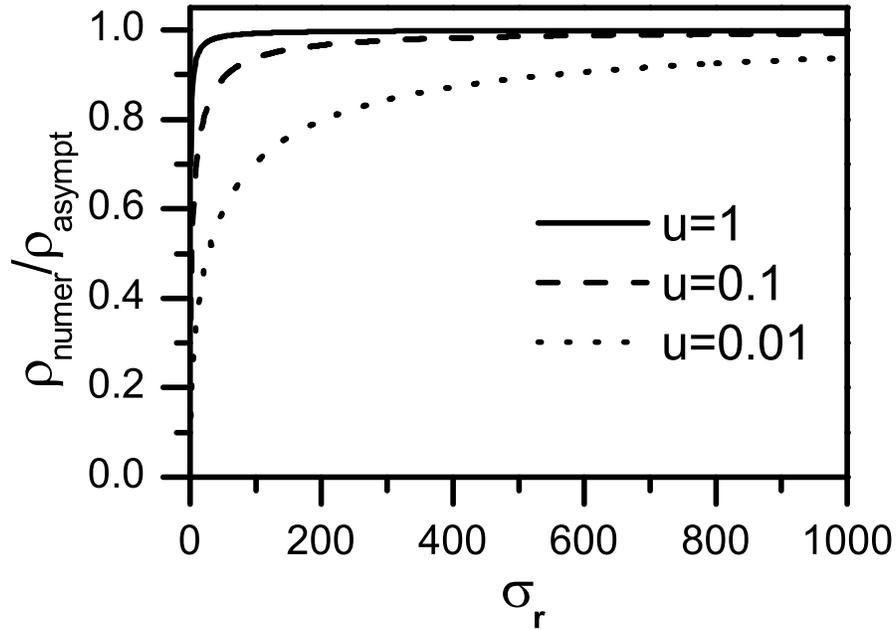}
\end{center}
\caption[]{The calculated numerically exactly function
${\cal {R}}_{cb}(\bm r)$, denoted $\rho_{numer}$, divided by its
asymptotics $\rho_{asympt}$ that up to a constant
is given by (\ref{ourasymp}), versus distance $\sigma_{\bm r}$,
for $D=2$, $\chi =1$ and three values of $u$.}
\label{fig3}
\end{figure}

To obtain an asymptotics stronger than the TDE asymptotics, valid for small and
large values of $u$, we  have applied to the integrals in (\ref{Rcb}) the Laplace
asymptotic expansion. Unfortunately, we have succeeded to carry out the
calculations only in the $D=1$ case (more results on the correlation function
in this case can be found in \cite{jkd}) and for the diagonal direction in the $D=2$
case. The result is of the form (\ref{asymptot}), with the inverse correlation
length
\be
\label{ourasymp}
\xi^{-1}(\epsilon)=\ln(\epsilon+\sqrt{\epsilon^2+1}),
\ \ \ \ \epsilon =(2D)^{-1}u,
\ee
and the exponent $\gamma=D/2$. Making a similar comparison as that displayed in
Fig.~\ref{fig2}, we find a satisfactory agreement, see Fig.~\ref{fig3}.
Apparently, in the limit
of the vanishing gap the inverse correlation length (\ref{ourasymp}) vanishes
proportionally to the gap width. In the one-dimensional case, the same result
has been obtained numerically for periodic potentials of period higher than 2
\cite{jkd}.

\begin{figure}
\begin{center}
\includegraphics[clip=on,width=8.2cm]{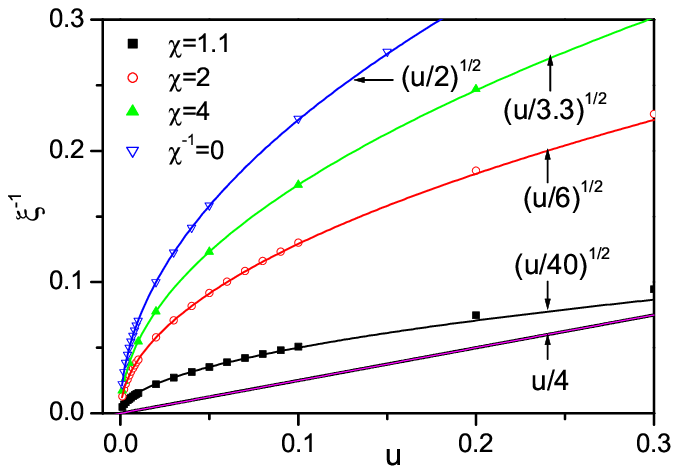}
\hfill
\includegraphics[clip=on,width=8cm]{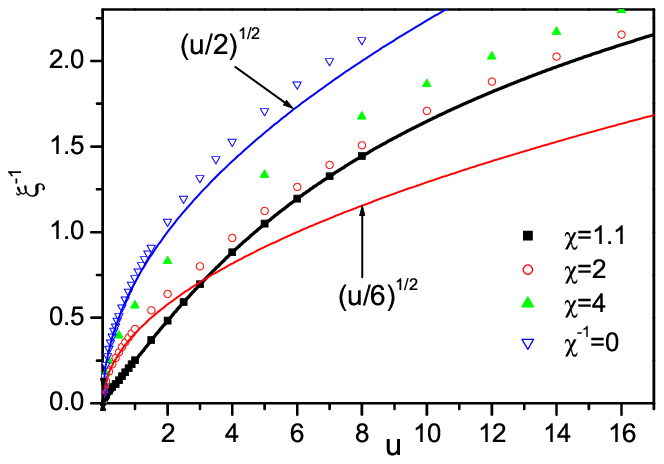}
\end{center}
\caption[]{The inverse correlation length, $\xi^{-1}$,
obtained from the calculated numerically exactly
function ${\cal {R}}_{cb}(\bm r)$, versus $u$,
for $D=2$ and three non-diagonal directions. Thin continuous curves
are the fits by the function $const \sqrt{u}$.
The thick curve represents $\xi^{-1}$ in the diagonal direction,
according to (\ref{ourasymp}).}
\label{fig4}
\end{figure}

Assuming that the asymptotic behavior of ${\cal {R}}_{cb}(\bm r)$ is of the
form (\ref{asymptot}) with the exponent $\gamma$ independent of $u$,
we have found numerically, see Fig.~\ref{fig4}., that in non-diagonal directions the
inverse correlation length vanishes as $\sqrt u$ as $u \rightarrow 0$.
For sufficiently large $u$, depending on the direction, the correlation
length behaves as in the diagonal case, i.e. obeys (\ref{ourasymp}).
Our data show also that the decay rate of DM elements is the slowest in the
diagonal direction.\\[0.5cm]

In summary, this work has been inspired by the recent paper of
Taraskin et al \cite{taraskin1}  on spatial decay of DM in a
translationally invariant model of insulator -- the TDE insulator.
Having found that the TDE model shows unphysical features, we have
carried out similar studies of another model of insulator -- the
chessboard insulator, whose translational symmetry is broken. In
contrast to TDE insulator, in the chessboard insulator there is a
unique energy parameter which determines both, the gap between the bands
and the DM elements. We have succeeded in deriving analytically
a dimension-dependent power-law factor, predicted in \cite{he},
with the exponent $\gamma = D/2$, and a gap-dependent exponential
decay rate, discussed extensively in \cite{beigi}.
Specifically, in $D=1,2,3$ cases,
the DM elements of the chessboard insulator obey the asymptotics
derived for TDE insulator by Taraskin et al, which we have found
to hold only for very large gap widths.
In $D=1$ case and along the diagonal in $D=2$ case, we
have derived a stronger asymptotic formula, valid for
arbitrary gap widths. The latter asymptotics implies that the
inverse correlation length of the chessboard insulator vanishes
linearly with the vanishing gap width. Concerning non-diagonal
directions, on assuming that the exponent $\gamma$ is independent
of the gap width, we have found numerically that the inverse
correlation length vanishes as the square root of the gap width.
Independently of the direction, the inverse correlation length
grows logarithmically with the gap width, for sufficiently large
gaps. A more comprehensive discussion will be given in a separate
article \cite{jk}.

T.K. is grateful to the Institute of Theoretical Physics of the
University of Wroc\l aw for kind hospitality and financial support.

\end{document}